\documentclass[reprint,
superscriptaddress,
amsmath,amssymb,aps,showkeys,showpacs,
twoside,final,secnumarabic,%raggedbottom,
nofootinbib]{revtex4-2}

\usepackage[paperwidth=205mm,paperheight=290mm,top=17mm,bottom=25mm,
inner=17mm,outer=17mm,
twoside]{geometry}

\usepackage{cmap}
\usepackage[utf8]{inputenc}
\usepackage[english]{babel}
\usepackage{color}
\usepackage{graphicx}% Include figure files
\usepackage{dcolumn}% Align table columns on decimal point
\usepackage{bm} % bold math
\usepackage[unicode=true,colorlinks=true,linkcolor=magenta, urlcolor=blue, citecolor = blue,breaklinks]{hyperref}
\usepackage{multirow}
\usepackage{url}
\usepackage{breakurl}

\DeclareGraphicsExtensions{.eps}

\newcount\issue
\newcount\Vol
\newcount\numb
\usepackage{fancyhdr} %this packages %provides fancy up and bottom of page
\def\Vol{\textbf{78}}
\def\numb{x}
\begin{document}

%====== Начало шапки статьи  ============
\title{JOURNAL SECTION OR CONFERENCE SECTION\\[20pt]
Neutrino masses in cosmology} 

\def\addressa{Istituto Nazionale di Fisica
  Nucleare (INFN), Sezione di Torino, Via P. Giuria 1, I-10125 Turin,
  Italy}
\def\addressb{Instituto de Fisica Teorica, CSIC-UAM
C/ Nicolás Cabrera 13-15, Campus de Cantoblanco UAM, 28049 Madrid, Spain}

\author{\firstname{S.}~\surname{Gariazzo}}
\email[E-mail: ]{stefano.gariazzo@ift.csic.es}
\affiliation{\addressa}
\affiliation{\addressb}

\newcommand{\Neff}{\ensuremath{N_{\rm eff}}}
\newcommand{\mnu}{\ensuremath{\Sigma m_{\nu}}}
% \received{xx.xx.2023}
% \revised{xx.xx.2023}
% \accepted{xx.xx.2023}

\begin{abstract}
We review the status of neutrino mass constraints obtained from cosmological observations, with a particular focus on the results derived considering Cosmic Microwave Background (CMB) data by various experiments (Planck, ACT and SPT), Baryon Acoustic Oscillation (BAO) determinations and other late-universe probes. We discuss the role played by priors and parameterizations in the Bayesian analyses, both at the time of determining neutrino masses or their ordering, and compare cosmological bounds with terrestrial constraints on both quantities.
\end{abstract}

% \pacs{Suggested PACS}\par
% \keywords{Suggested keywords   \\[5pt]}
%DOI:  

\maketitle
\thispagestyle{fancy}

\section{Neutrino masses}\label{sec:numasses}
Within the Standard Model of Particle Physics, neutrinos are the only particles of which we still ignore the absolute mass.
Constraints on the effective neutrino mass by KATRIN \cite{KATRIN:2021uub},
through the observation of the end point of the beta-decay spectrum of tritium,
tell us that neutrinos are at least six orders of magnitudes lighter than the lightest charged fermion, the electron.
Because of this huge scale difference, neutrinos were believed to be massless for a long time, and indeed in the original formulation of the Standard Model, neutrinos are the only fermions with zero mass.
In the standard model, neutrino flavors are defined by their interaction with charged fermions.
For the first time more than 20 years ago \cite{Fukuda:1998mi,Ahmad:2001an,Ahmad:2002jz},
experiments provided evidence that
neutrinos with a given flavor can oscillate to a different flavor while propagating.
This phenomenon, neutrino oscillations,
implies that neutrinos exist in two different basis, one describing their propagation (mass basis) and one the interaction (flavor basis).
Neutrino oscillations
guarantee us that at least two of the three neutrino states must have a positive mass:
the change of flavor cannot take place if any of the two neutrino eigenstates have the same mass.
Neutrino oscillation experiments allow us to measure the two squared-mass differences,
the three mixing angles and the CP violating phase
that describe the mixing between flavor and mass eigenstates,
see e.g.~\cite{Gonzalez-Garcia:2021dve,Capozzi:2021fjo,Esteban:2020cvm,deSalas:2020pgw}.

Although one of the two mass splittings, mostly related to solar neutrino oscillations, is known to be positive,
the sign of the second one, mostly associated with atmospheric neutrino oscillations,
has not been determined yet.
Because of this, we still have two possibilities for the ordering of neutrino masses, see e.g.~\cite{deSalas:2018bym}:
it may be \emph{normal} if the atmospheric mass splitting is positive and the lightest neutrino is the one with the largest electron flavor component,
or \emph{inverted} if the mass splitting is negative and the lightest neutrino has the smallest mixing with the electron flavor.
Given the values of the two mass splittings~\cite{Gonzalez-Garcia:2021dve,Capozzi:2021fjo,Esteban:2020cvm,deSalas:2020pgw},
we know that the second-to-lightest neutrino must have a mass larger than 0.009~eV if the ordering is normal and 0.05~eV in the inverted case.
The third mass splitting, in turn, must have a mass larger than approximately 0.05~eV in both orderings.
Thanks to these considerations, we can determine the minimum value for the sum of the three neutrino masses, \mnu, which cannot be lower than $\sim0.06$~eV in normal ordering or $\sim0.1$~eV if the ordering is inverted.
Because of these lower limits being different, cosmological measurements of neutrino masses could help to determine the neutrino mass ordering, as we will discuss in the following.
Before addressing the neutrino mass ordering determinations, however,
let us focus on how cosmology can constrain the neutrino mass scale.

\section{Neutrino masses in cosmology}
In cosmology, neutrinos play different roles depending on the epoch one considers.
In the early universe, neutrinos were relativistic and they contributed to the total amount of radiation.
It is common to describe their amount in terms of the effective number of relativistic degrees of freedom, \Neff, which quantifies the ratio between neutrino and photon energy densities.
The most precise theoretical determinations report $\Neff=3.044$ for active neutrinos \cite{Froustey:2021azz,Bennett:2020zkv,Froustey:2020mcq,Akita:2020szl}
(see also \cite{Cielo:2023bqp}),
in perfect agreement with Planck measurements of the Cosmic Microwave Background (CMB) radiation \cite{Planck:2018vyg}.
After their decoupling, relic neutrinos propagate freely in the universe and their temperature is redshifted until today, when it is expected to be around $10^{-4}$~eV.
Since at least two of the three neutrinos have a mass higher than this value,
relic neutrinos become non-relativistic at some point during the universe expansion.
In the late universe, therefore, neutrinos contribute as matter and their energy density is proportional to their mass multiplied by the number density.
Because of this relation, if we assume stable neutrinos with a fixed mass and a standard decoupling scenario, we can constrain their mass by measuring their energy density today.

Several cosmological observables are sensitive to the presence of neutrinos and provide us a way to constrain their energy density, see e.g.~\cite{Lesgourgues:2018ncw}.
On the one hand, massive neutrinos affect the late-time evolution of the CMB spectrum,
which has been observed with outstanding precision by the Planck satellite.
The impact of neutrino masses on the CMB, however, is not very strong and is degenerate with the effect of other cosmological parameters such as the Hubble factor $H_0$.
As a consequence, the CMB alone cannot put the strongest constraints on the sum of the neutrino masses.

The transition from relativistic to non-relativistic particles, however, affects the formation of structures, which is much more relevant and enables to put much stronger constraints on \mnu.
Relativistic particles cannot be trapped by small gravitational potential wells and their pressure has the effect pushing matter out of small structures, hence they reduce the formation of structures at very small scales: they free-stream because of their velocity.
On the contrary, non-relativistic particles cluster because of the effect of gravity.
The transition from relativistic to non-relativistic, therefore, generates a change in behavior that leaves an imprint on the formation of structures.
Such imprint can be detected by looking at the matter power spectrum, at a specific scale which depends on the mass of the particle.
In terms of percentage, variations in the matter power spectrum due to the presence of massive neutrinos
can be a factor 10 larger that those on the CMB spectrum, hence why neutrino masses can be measured in much more details when the former is considered.
According to forecasts, ongoing matter power spectrum observations should enable us to reach a sensitivity of $\approx0.02-0.03$~eV on the sum of the neutrino masses, see e.g.~\cite{Audren:2012vy,Font-Ribera:2013rwa},
while future experiments can push this value even further, see e.g.~\cite{Brinckmann:2018owf}.
Even with such precision, however, it seems unlikely that cosmology will be able to test individual neutrino masses \cite{Archidiacono:2020dvx}.
In this sense, cosmology may help in providing information about the neutrino mass ordering
only by testing whether \mnu\ is smaller than the minimum value of 0.1~eV allowed for inverted ordering, as we will discuss later,
but it will not determine the mass ordering directly by the differences between the three neutrino masses.

Cosmological constraints on \mnu\ are, in any case, indirect probes on the energy density of neutrinos, subject to variations according to the model one considers for describing the evolution of the universe.
The basic model, which we call $\Lambda$CDM after the two main components of the energy density today
(the cosmological constant $\Lambda$ and Cold Dark Matter), is based on the assumptions that the universe
is approximately homogeneous and isotropic at large scales, and that General Relativity holds.
The minimal amount of parameters one needs to describe such model are six:
the amount of dark matter and baryon energy densities,
the optical depth to reionization,
the Hubble factor today $H_0$ (or equivalently the angular scale of CMB peaks),
the amplitude and tilt of the power spectrum of initial curvature fluctuations.
As we will discuss later, most of the constraints are obtained by considering minimal variations of this basic model, therefore leaving out any other possible deviations from the standard case.
Additional freedom in the theoretical model, in turn, may correspond to less constraining bounds because of the degeneracies between \mnu\ and the new parameters.

\section{Constraining neutrino masses with cosmology}

If we only consider CMB measurements, within the minimal 7-parameters model named $\Lambda$CDM+\mnu, the best determination for the sum of the neutrino masses is $\mnu<0.26$~eV when ignoring CMB lensing constraints and $\mnu<0.24$ when they are included \cite{Planck:2018vyg},
both at 95\% confidence level (CL).
The bound is already much stronger than the ones obtained with pre-Planck CMB observations,
thanks to the complementarity between temperature and polarization measurements.
When low-redshift probes are considered, the bound improves as expected.
The Planck collaboration reports $\mnu<0.12$~eV at 95\% CL when including Baryon Acoustic Oscillations
as reported in the Data Release 12 (DR12) \cite{Alam:2016hwk} of the BOSS experiment \cite{Dawson:2012va}.
Notice that the value of the upper \mnu\ bound is close to the minimum value allowed for inverted ordering, 0.1~eV.

More recently, the BOSS collaboration released new observations, which include the full dataset denoted as DR16 \cite{Alam:2020sor}.
With these new data, the upper bound on \mnu\ decreases,
reaching values below the minimum allowed for inverted ordering:
$\mnu<0.09$~eV \cite{DiValentino:2021hoh} when considering a combination of CMB and lensing observations by Planck, supernovae as reported in the Pantheon compilation \cite{Scolnic:2018rjj} and a combination of DR12 and DR16 data by BOSS.
As we will discuss in the incoming section, such result can start to seriously question the inverted ordering scenario.

Among particle physicists, one of the strongest critiques against cosmological constraints on neutrino masses is that the results depend on the assumptions on the model and parameter priors, and as a consequence, limits can vary a lot.
In the past, indeed, it was easy to relax the bounds on \mnu\ by leaving one or two additional cosmological parameters as free in the analysis, see e.g.~\cite{Planck:2013pxb},
which reports almost a factor two increase in the upper limit when the curvature of the universe is allowed to be different from zero
(at 95\% CL, we have $\mnu<0.25$~eV versus $\mnu<0.44$~eV, respectively for fixed or free curvature, when considering the first Planck CMB public dataset together with available BAO measurements at the time),
even when BAO data are taken into account.
Nowadays, although the bounds are relaxed when additional parameters are let free,
the difference is not so significant.
If we consider the same example mentioned above, the limit goes from $0.09$~eV (flat universe)
to $0.11$~eV (free curvature),
always at 95\% CL, when considering the most constraining dataset available today
(Planck measurements, supernovae by the Pantheon compilation and the BOSS DR16 results).
Other extensions of the minimal $\Lambda$CDM+\mnu\ model correspond to similar behavior,
although multiple additional free parameters can still give the flexibility to relax the limit by a significant factor.

One can notice, however, that multiple free parameters are not justified at the time of considering the theoretical model, if the quality of the fit does not improve in a significant way.
According to Occam's razor, indeed, a model is better than another when it can reproduce data equivalently well, but with a smaller number of parameters.
Bayesian statistics allows to incorporate the Occam's razor easily when performing model comparison by means of the Bayesian evidence and the Bayes factor, see e.g.~\cite{Trotta:2008qt}.
As a derivative of Bayesian model comparison,
it is also possible to perform a marginalization over different models,
in order to obtain more robust limits on a single parameter that take into account
the possible existence of different theoretical scenarios \cite{Gariazzo:2018meg}.
In the context of cosmological neutrino mass bounds, the model marginalization has been proposed for example in \cite{diValentino:2022njd},
where the most constraining dataset available today has been considered.
It is interesting to see that, when considering a set of possible one-parameter or two-parameter extensions of the $\Lambda$CDM+\mnu\ model,
the limit grows from the most constraining 0.09~eV of the minimal scenario to approximately 0.1~eV (95\% CL) for the model-marginalized case.
As a consequence, we can say that the limit is sufficiently robust given the most recent available observations.
We can also notice that switching from Planck CMB measurements
to other CMB probes, such as a combination of WMAP~\cite{Hinshaw:2012aka,Bennett:2012zja} and ACT~\cite{Choi:2020ccd,Aiola:2020azj} or SPT-3G~\cite{SPT-3G:2021eoc} data,
does relax the obtained model-marginalized limits on \mnu\ to some value between 0.25~eV and 0.3~eV (95\% CL)  \cite{DiValentino:2023fei},
which are still stronger than terrestrial probes of neutrino masses.

Currently, there is also another problem related to the Bayesian calculation of neutrino mass limits from cosmology.
Cosmological constraints are always computed through Bayesian methods.
According to the Bayes theorem, the posterior probability distribution
(describing our knowledge about a certain set of parameters after we performed an experiment)
is proportional to the prior probability distribution (what we knew before performing the experiment)
multiplied by some likelihood function, that describes the outcome of the experiment.
In the specific case of neutrino masses, the prior probability distribution, or simply ``prior'',
is certainly zero for negative masses, but its shape in the positive range can be decided by the researcher performing the analysis.
There are theoretical arguments that motivate the prior probability to be constant either
in the neutrino mass or in its logarithm, for example,
but there is no good justification for the limits one considers for the prior.
Notice that the numerical value of the limit, for example at 95\% CL, requires to find the region of parameter space which encloses 95\% of the posterior probability through an integral.
If one considers a uniform prior in $\log\mnu$, for example,
the integral of the posterior probability depends on the lower allowed value for \mnu\ and changing the prior limit may disrupt the obtained results.
This problem is a purely numerical volume effect that arises every time the likelihood is not able
to constrain the parameter on both ends on the prior: we talk in this case of a \emph{open} likelihood.
How to compute constraints on parameters for which the likelihood is open has been argument of theoretical discussion in Bayesian context for a long time.
For this reason, methods have been developed in order to avoid
presenting bounds that require to integrate the posterior distribution,
but rather study how the likelihood behaves.
One noticeable method, called relative belief updating ratio \cite{DAgostini:2000edp,Gariazzo:2019xhx},
requires to compare the Bayesian evidences computed at two fixed values of the parameter under consideration,
and use the obtained quantity in the same way one would use for Bayesian model comparison,
instead of integrating the posterior probability to compute credible intervals.
In simple cases, this method is equivalent to a profile likelihood study,
but it can catch effects related to the bulk of the posterior distribution and marginalize over the rest of the varying parameters instead of considering only one point of the parameter space at each time.
Moreover, it can be computed directly from the Bayesian prior and posterior distributions
(obtained from Markov Chain Monte Carlo or nested sampling algorithms, for example),
without performing a time-consuming maximization of the likelihood.
In our case, the relative belief updating ratio presents as a flat function for low neutrino masses (where cosmological data cannot distinguish the presence of massive neutrinos)
and rapidly decreases at high mass values,
when data effectively disfavor the presence of a non-relativistic neutrino component.
We believe that presenting \mnu\ limits according to this function may help to reduce the subjectivity in the computation, in particular regarding the lower limit of the \mnu\ prior.

\section{Constraining the mass ordering}
As already mentioned,
if one takes into account the information on squared-mass differences provided by neutrino oscillation experiments,
the minimum value for the sum of neutrino masses is $\approx0.06$~eV in normal ordering and 0.1~eV in the inverted ordering case.
If cosmology disfavors values above 0.1~eV, in practice it disfavors inverted ordering
(under the assumptions of stable neutrinos, standard cosmological evolution and standard correspondence between neutrino masses and non-relativistic neutrino energy density).
Claiming that an upper limit such as $\mnu<0.1$~eV at 68\% CL excludes inverted ordering at 68\% CL, however, is inaccurate,
because part of the normal ordering parameter space is also excluded:
the comparison among the two mass orderings must be performed by means of model comparison techniques~%
\footnote{Notice that we denote here a ``model'' in a Bayesian statistic sense:
a ``model'' is defined by a set of parameters that describe the theory and their prior probability,
so that it can also be indicated as ``parameterization''.
The underlying theoretical model (the standard model of particle physics, for example) is the same,
but we can study the same theoretical model with infinite different parameterizations.
Parameterizations are not equivalent under the Bayesian framework
because the volume of the parameter space changes depending on which parameters and priors we choose.}.
In this case, we have two competing and mutually exclusive parameterizations.
In a Bayesian context, the probability in favor of one of the two can simply be obtained by comparing the Bayesian evidences of the ``normal ordering'' (NO) versus the ``inverted ordering'' (IO) scenarios against a chosen dataset,
which may include combinations of cosmological and terrestrial data.
Bayesian evidences measure how well the model describes the data over the entire parameter space, through an integral of the likelihood over the entire multi-dimensional prior.
The only difference between the two parameterizations we are interested in is the sign of one of the two mass splittings: the rest of the parameters are exactly the same, hence the volume of the parameter space is the same.
The question, however, is: how do we parameterize the three neutrino masses in the optimal way?

The simplest thing to do from a cosmological perspective is probably to consider the three neutrino masses $m_i$ as independent parameters, as cosmology directly measures the mass scales.
On the other hand, oscillation experiments test directly the mass splittings ($\Delta m^2_{ij}=m^2_i-m^2_j$), not the absolute neutrino masses,
so that another possible choice is to consider two mass splittings plus one absolute mass scale:
the latter can either be the mass of the lightest neutrino or the sum of the three masses.
Moreover, since we ignore the absolute value of the neutrino mass scale, a uniform prior
on $\log m_i$ seems more reasonable in order to give all masses the same initial probability,
but $\mnu$ is constrained to a rather small range and a uniform prior on \mnu\ is perfectly justified.
All these possibilities have been considered in the past, see e.g.~\cite{Caldwell:2017mqu,Gariazzo:2018pei,Gariazzo:2022ahe},
with rather controversial conclusions about the mass ordering,
due to various subtleties hidden in Bayesian model comparison methods,
which we discuss now.

As we said, Bayesian model comparison is a matter of confronting two parameterizations
and how well they perform in the parameter space,
by means of the Bayes factor, which is a ratio of the Bayesian evidences of the two competing models.
In principle, only experimental data should be responsible of the obtained result.
It is possible to tweak the results by changing the shape and range of the prior, though:
because of the automatically incorporated Occam's razor, a parameterization with a very wide prior is disfavored over one which only allows the parameters to vary close to the best-fit region.
In order to have robust conclusions, therefore, it is important to check whether the choosen parameterization does not artificially affect the results.
The first crucial test when one wants to understand how the parameterization affects the mass ordering determination
is to study the Bayes factor in a simple case where the considered data sets do not have a preference in favor of any of the mass ordering.
For instance, if we only consider the neutrino oscillation constraints on the absolute value of the mass splittings and the same priors and ranges for the two parameterizations,
we should obtain that the two competing options are equivalent,
because the parameter space is the same and there is no preference for any mass ordering in the data.
Even in this simple case, however, it turns out that the parameterization with three neutrino masses has a preference in favor of normal ordering if the prior is uniform on the logarithm of the neutrino masses \cite{Gariazzo:2018pei}.
The reason is related to the fact that a significantly larger part of the inverted ordering parameter space
($m_2<0.05$~eV for inverted ordering versus $m_2<0.01$~eV in normal ordering)
is excluded by data.
While this difference is almost irrelevant when considering a uniform prior on $m_i$,
the logarithm amplifies the disparity to a significant level.
This is therefore an example of an \emph{artificial} preference in favor of normal ordering,
which has no origin in data but rather on personal choices of the prior.
Since this effect does not appear when considering a parameterization with one absolute mass scale (either \mnu\ or the lightest neutrino mass are equivalent)
plus two mass splittings,
neither using uniform priors or logarithmic ones,
it is recommended to stick to this latter choice of parameters in order to avoid
unwanted biased results \cite{Gariazzo:2018pei,Gariazzo:2022ahe}.

Neutrino oscillations nowadays have a small preference (approximately 2$\sigma$) in favor of normal ordering, see e.g.~\cite{Gonzalez-Garcia:2021dve,Capozzi:2021fjo,Esteban:2020cvm,deSalas:2020pgw}.
From cosmology alone, unsurprisingly, we obtain the maximum preference in favor of normal ordering when considering BOSS DR16 data in combination with Planck CMB measurements and Pantheon supernovae observations,
which corresponds to the strong limit $\mnu<0.09$~eV at 95\% \cite{DiValentino:2021hoh}.
Such preference is at the level of $1.4\sigma$ \cite{Gariazzo:2022ahe}.
The combination of cosmological and terrestrial observations yields a significance around $2.7\sigma$ when conservative model and prior choices are assumed \cite{Gariazzo:2022ahe}.
Given current data, stronger claims are always due to artificial volume effects arising from the selected parameterization.

Finally, let us now try to answer the following question: will cosmology alone be able to disfavor inverted ordering and actually confirm normal ordering to a high significance?
Related to this point, let us recall that the incoming generation of cosmological probes
is expected to reach a precision at the level of 0.02~eV on \mnu~\cite{Brinckmann:2018owf},
when considering for example a combination of Planck data plus Euclid observations.
The sensitivity is a factor two better when considering future CMB experiments,
or even more when SKA data will be available.
In this case, let us simply consider the target for the incoming generation, $\sigma(\mnu)\simeq0.02$~eV.
It remains clear that any true value of \mnu\ equal or greater than 0.1~eV does not allow to determine the mass ordering, because such value would be allowed both in the normal and inverted ordering cases.
With such level of precision, cosmology will reach a preference of approximately $2\sigma$ in favor of NO, if this is the correct mass ordering and the true value for the sum of neutrino masses is the minimal 0.06~eV \cite{Gariazzo:2023joe}.
This is expected, in the sense that a measurement $\mnu\approx0.06\pm0.02$~eV excludes the value 0.1~eV at approximately $2\sigma$ significance.
Current cosmological measurements, however,
have a preference for $\mnu\approx0$ (unphysical in light of neutrino oscillations),
mostly driven by Planck.
If future experiment will still confirm such preferred value,
then normal ordering might be favored at a higher level of precision with respect to inverted ordering (up to $\approx4\sigma$),
but this result would actually correspond to a strong tension between terrestrial and cosmological observations.
There are several possible methods that allow to quantify the level of tension,
both Bayesian and frequentist (see e.g.~\cite{Gariazzo:2023joe}).
For a measurement $\mnu\approx0.00\pm0.02$~eV from cosmology, the tension with terrestrial observations would be
at the $3\sigma$ level in normal ordering and $>4\sigma$ for inverted ordering,
with small variations depending on the selected statistical test.
Notice that such a tension would imply that something is wrong with our assumptions about the neutrino behavior in cosmology or the cosmological model itself.
Possible solutions to this problem (decaying neutrinos, time-varying neutrino masses and so on)
are outside the scope of the present discussion.

\section{Conclusions}
Neutrinos are the only particle in the standard model of particle physics of which
we still ignore the absolute mass scale.
Neutrino masses are constrained by terrestrial experiments to be at least six orders of magnitude smaller than those of the lightest elementary charged fermion, the electron.
The existence of neutrino oscillations, however, tells us that at least two neutrino are massive and experimental data let us put a lower bound on two of the three masses.
Such lower limits impose that the sum of the neutrino masses cannot be smaller than a specific value.
Given the fact that we have two possibilities for the neutrino mass ordering (normal if the lighest neutrino has the largest mixing with the electron flavor, inverted if the mixing is the smallest),
the smallest possible value of the sum of neutrino masses is $\approx0.06$~eV or 0.1~eV, respectively for normal and inverted ordering.

Cosmology can help us to constrain the sum of neutrino masses indirectly, thanks to the fact that massive neutrinos
behave as non-relativistic particles at late times in the universe history,
and their energy density is proportional to their mass.
Assuming the standard cosmological model in its minimal form, stable neutrinos and the standard conversion factor between energy density and neutrino mass,
the most constraining cosmological bound on the sum of neutrino masses is currently
$\mnu<0.09$~eV at 95\% CL \cite{DiValentino:2021hoh}.
If one takes into account additional freedom in the cosmological model (e.g.~free curvature, additional relativistic species, non-standard dark energy equation of state)
and marginalizes over a set of possibilities,
this limit is relaxed to $\mnu<0.1~eV$ (95\% CL) given the same dataset \cite{diValentino:2022njd}.
This states the fact that neutrino mass bounds from cosmology are quite reliable and robust to simple variations on the cosmological model.
Given the expected sensitivity at the level of 0.02~eV on \mnu\
which might be achieved by combining current CMB data from Planck
and low-redshift observations by the ongoing experiment Euclid,
we can expect incoming observations to provide a first cosmological determination of \mnu.

The strongest cosmological limits on \mnu\ start to question the possibility that the neutrino mass ordering is inverted,
as values $\mnu>0.1$~eV are disfavored by data.
Bayesian model comparison allows to quantify the preference in favor of normal ordering,
which is currently at the level of $1.4\sigma$ from cosmology alone, $\sim2\sigma$ from oscillation experiments alone, and $2.7\sigma$ when cosmology and terrestrial experiments are combined.
One must be careful when using Bayesian statistics for performing mass ordering studies,
because of the possible artificial strong preference in favor of normal ordering
that may arise in case specific choices are made regarding the choice of parameters and priors for the analysis.
The sensitivity to the mass ordering from incoming cosmological experiments is not expected to increase significantly, because
cosmology can only determine the mass ordering to a high significance if it is normal,
the lightest neutrino is almost massless and the precision on the \mnu\ value increases to at least $\sigma(\mnu)\approx0.01$~eV.
An error of 0.02~eV on \mnu\ by Planck CMB constraints plus low-redshift probes by Euclid,
might correspond to a $\gtrsim2\sigma$ preference in favor of normal ordering from cosmology alone,
depending on the true value of \mnu.
However, current cosmological measurements (especially from Planck)
seem to prefer $\mnu\approx0$, unphysical if one takes into account the information on the mass splittings we obtain from neutrino oscillations.
If the current preferred value will be confirmed by future probes,
then we might obtain a strong preference in favor of normal ordering,
originated however by a strong tension between cosmological and terrestrial data.
In such case, rather than determining the mass ordering, we will discover
that some big change is required in how neutrino behave in the universe or in the way we describe the evolution of the universe.

\begin{acknowledgments}
S.G.\ is supported by the European Union’s Framework Programme for Research
and Innovation Horizon 2020 (2014–2020) under grant agreement 754496
(FELLINI, until September 2023) and Junior Leader Fellowship LCF/BQ/PI23/11970034 by La Caixa Foundation (from October 2023).
\end{acknowledgments}

%%%%%%%%%%%%%%%%%%%%%%%%%%%%%%%%
% USE thebibliography
%%%%%%%%%%%%%%%%%%%%%%%%%%%%%%%%
\bibliography{main.bib}

\end{document}